RUSSIAN ACADEMY OF SCIENCES
PETERSBURG NUCLEAR PHYSICS INSTITUTE

G. I. Shulyak, A. A. Rodionov

# Programs for the work with ENSDF format files: Evaluator's editor EVE, Viewer for the nuclear level schemes

**A b s t r a c t**

Tools for the regular work of the nuclear data evaluator are presented: the context-dependent editor EVE and the viewer for the level schemes of nuclei from ENSDF datasets. These programs may be used by everybody who works with the Evaluated Nuclear Structure Data File and for the educational purposes.

## Introduction

The Evaluated Nuclear Structure Data File (ENSDF) is the basis of a large number of databases on ground and excited nuclear states [1]. That file is supported in the actual state by the International Network of Nuclear Structure and Decay Data Evaluators [2]. The Network consists of more than 20 regional Centers situated in different countries over the world. Each Center carries out the work on the analysis and evaluation of the published experimental information on nuclear states and transitions. The result of the work is the text file written in the ENSDF format. The ENSDF file as a whole is being completed and saved by efforts of the National Nuclear Data Center (NNDC), Brookhaven National Laboratory, USA. The NNDC BNL group prepares for publication the periodic journal *Nuclear Data Sheets* which is totally based on the information from the ENSDF. The syntax of the ENSDF was first proposed at the Nuclear Data Project, Oak Ridge National Laboratory [3]; the current format is described in [4].

The present article describes programs Evaluator's Editor EVE and data_set_viewer which may be used in the regular work of the evaluator of nuclear data. These programs may be used separately and with the program package ENSDF_toolbox [5] also. Programs EVE and data_set_viewer are written in C language and operate in the Linux system. The EVE program may be linked as the version for text terminal (on the base of the ncurses library; accompanying libraries are form, menu) and for X window graphic environment (gdk, glib, gnome, gtk+ libraries). The data_set_viewer program operates in the X window graphic environment and uses gnome and gtk+ libraries.

## The Evaluator's Editor EVE

The significant part of the work of an evaluator of nuclear data is the editing and complementation of the ENSDF. Different text editors of general purpose are used for this purpose. Here we propose to use for this work the special context-oriented editor EVE (EValuator's Editor) which makes this process easier. This editor checks the data for consistency with the ENSDF syntax [4] at input directly and may check existing datasets in addition to other checkers. Evaluators use the FMTCHK and PANDORA programs for checking of their files to the ENSDF format and content (in part). These programs may be loaded from the NNDC BNL site http://www.nndc.bnl.gov. Nevertheless in practice one can find a significant number of misprints in the ENSDF, about 10,000 to the spring of 2009 y. The usage of the EVE program should allow to reduce this number.

The working area of the editor EVE consists of fields marked by different colors. It makes the structure of the ENSDF datasets demonstrative. In addition, the program analyses "on the fly" whether the data are accordant to the ENSDF format.

The editor EVE has an opportunity of tuning of its properties such as the structure of the working area, rules for checking and others. These properties are described in the configuration files which have the text format and may be corrected if necessary. After some adjustment the EVE program may be adapted for the work with other formats, not for the work with ENSDF only.

### *Start of the editor*

Editor may be started from the command line by the command:

```
eve
```

or

```
eve file_name
```

for the editing of any ENSDF file.

For viewing of the file the command is:

```
eve -r file_name
```

### *The screen of the editor*

The screen of the editor (fig.1) consists of 3 parts.

1) The top line of the screen is the status line:

| 210 | (8594) | [- - - O] | REFERENCE | .IDENT | /tmp/ENSDF_selection.FiTZde |
|---|---|---|---|---|---|
| current record | number of records | flags | name of record | name of field | filename, data from/to |

Flags are:

a) Flags of marking:

"-" means nothing is marked,
"F" means a field is marked,
"b" means the block of records is being marked,
"B" means the block of records is marked.

b) The flag of modification:

"-" means that the data were not modified from the moment of the loading or the last saving in a file by the key F2 or Shift+F2,
"M" means that there were changes in the data.

c) Reserved.

d) The flag of the mode of the replacement of symbols "O", or of the insert "-".

2) The working area.

3) The line at the bottom describes the action of the functional keys.

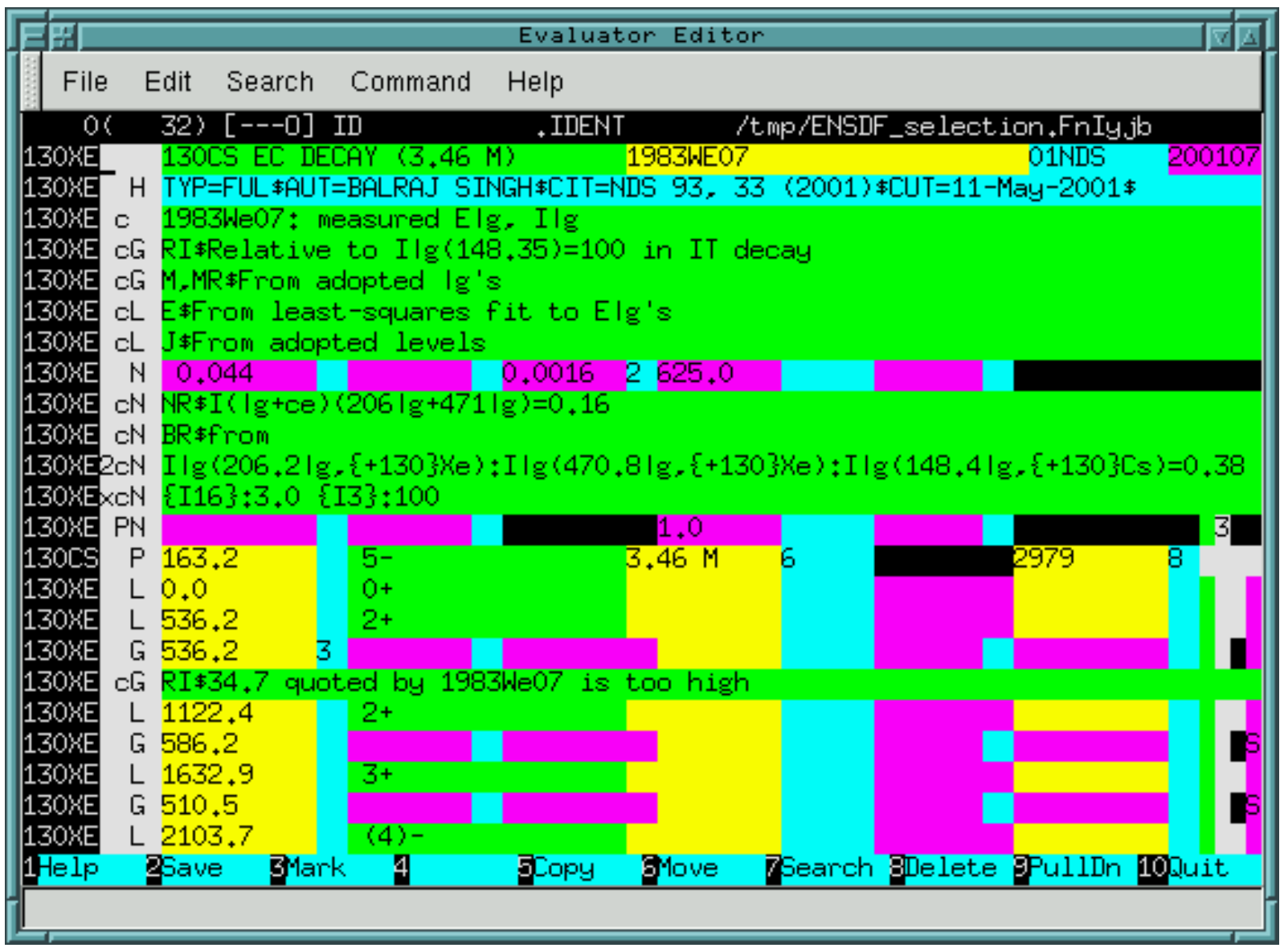

Fig. 1

***Keys***

*Keys for the editing of a field*

1) Arrows to the left/ right move the cursor to the left/right to one position, but within the given field only.

2) Home and End keys cause the jump of the cursor to the beginning or the end of a field, respectively.

3) Delete key deletes a symbol on a position of the cursor; Backspace deletes a symbol to the left of the cursor.

4) Insert key switches modes of the overlapping of symbols and the insertion of a symbol under the cursor. The current state of the key is indicated in the status line.

5) F8 acts depending on whether there is the marked field or the block. If none is marked, the current field is erased. If the field is marked, it will be erased. If the block of records is marked, it will be deleted.

*Keys for moving between fields/records*

1) TAB makes the next field of record to be current. If that field is the last in the current record, the first field of the next record becomes the current one. TAB+Alt makes the same but in the opposite direction.

2) The key Enter acts as the key TAB.

3) Arrows Up/Down move the cursor to one record upwards/downwards. If the last field is not NUCID or IDENT, the cursor moves to the field IDENT.

4) PageUp/PageDown keys move the cursor to the previous/next page. The size of the page is equal to the size of the working area minus one line.

5) F7 activates the search of the sample in the record. This key opens the window of the choice for the type of the record and for the input of a sample. In the field of the choice of the type of the record the choice may be made using Arrows Up/Down keys; a sample should be printed with the keyboard (expanded regular expressions [6] are admissible). Key Enter and the button "OK" initialize the search; the button "Cancel" interrupts the search.

6) The combination Shift+F7 initializes the search of the record. It opens the window of a choice of the type of records. The name of the record may be chosen by Up/Down keys. Key Enter initializes the search; the button "Cancel" interrupts the search.

7) Alt+E initializes the search of the next incorrect field. The cursor will be placed into that field.

8) Alt+L moves the cursor into the record with the defined number.

9) Alt+PageUp/Alt+PageDown moves the cursor to the top/end of the file.

*Operations with records and fields*

1) Shift+F3 marks the current field.

2) F3 switches on/off the mode of marking of records. The marked block of records may be moved (F6), removed (F8) or copied (F5).

3) F5 copies the content of the marked field to the current field. It may be used only in the cases when copied data and the destination field have the same format, that is, it is possible to copy values of spin/parity of record LEVEL to a fields of spin/parity LEVEL records only. The only exception is: the field NUCID of any record may be copied to the field NUCID of any record. If the block of records is marked, it will be copied before the current record.

4) F6 acts only, if the block of records is marked. The marked block will be deleted from its place and inserted before the current record.

5) F8 deletes the content of the current field. Shift+F8 deletes the current record.

6) Alt+D is the same as Shift+F8.

7) Alt+I inserts a record before the current record. If the field NUCID above or below the new record is not empty, it is copied in the field NUCID of inserted record. The inadmissible combination of symbols will be inserted in the field IDENT which does not allow movement to other fields until the correction of this field is made. After necessary correction all the record will have corresponding attributes.

### *Other keys*

1) F1 for help.

2) F2 saves the data in the current file. Combination Shift+F2 allows the change of the current name of the file.

3) Shift+F5 allows to insert the ENSDF formatted file the before the current record.

4) Ctrl+F saves the marked block in a file.

5) Alt+G allows to declare the field as “not erroneous”. It is accessible for any field except for NUCID and IDENT.

6) F9 calls the horizontal menu. The menu contains next items: “File”, “Edit”, “Search”, “Command”, and “Help”. Arrows Left/Right allow to choose the item, Enter opens it, F10 cancels it.

7) F10 closes the current session with the editor and saves the current content in the file ENSDF.ens.

8) Shift+F10 types the message about the last error found.

9) Ctrl+R calls re-drawing of the screen of editor. It may be useful if other task sends a message.

10) Alt+V calls the external program for viewing the level scheme of the marked block (only in the graphic X Window shell).

11) Alt+W calls the external program for viewing the level scheme of the data sets opened in the editor (only in the graphic X Window shell).

12) Ctrl+U realizes the undo function. Undo has the deep up to the beginning of editing or up to the last saving of the file by F2 or Shift+F2 key.

### *Auxiliary files*

Parameters and properties of the EVE editor are defined by next 3 auxiliary text files:

1) The file ENSDF.def describes all types of records and fields adopted in the ENSDF syntax. Determinations for all fields except NUCID (positions 1-5) and IDENT (positions 6-9) may be redefined. Here the example of the description for the LEVEL record is presented:

```
LEVEL name="The Level Record" regexp="^[ 12][ 0-9][0-9][A-Z0-9][ A-Z0-9]  L "
  .E     bounds=10:19 color=6 format=V.18 align=no
                      convert=upper name="Level energy"
  .DE    bounds=20:21 color=7 format=V.11 align=no
                      convert=upper name="Uncertainty E"
  .J     bounds=22:39 color=5 format=V.20 align=no
                      convert=upper name="Spin and parity"
  .T     bounds=40:49 color=6 format=V.14 align=no
                      convert=upper name="Half-life"
  .DT    bounds=50:55 color=7 format=V.12 align=no
                      convert=upper name="Uncertainty in T"
  .L     bounds=56:64 color=8 format=V.22 align=no
                      convert=upper name="Angular…"
  .S     bounds=65:74 color=6 format=V.21 align=no
                      convert=upper name="Spectroscopic…"
  .DS    bounds=75:76 color=7 format=V.11 align=no
                      convert=upper name="Uncertainty S"
  .C     bounds=77:77 color=5 format=V.8
                      convert=no    name="Comment FLAG"
  .MS    bounds=78:79 color=1 format=V.17 align=left
                      convert=upper name="Metastable..."
  .Q     bounds=80:80 color=8
                      convert=upper name="…"
```

Keywords for the description of a record and a field are:

**bounds** (boundaries of a field).
**color** (index of the color of a field; value is used in files GCOLORS.def, COLORS.def for the determination of the screen color).
**format** (description of the format from the correspondent paragraph in [4]).
**align** (disposition of the value: no, left, right).
**convert** (converting of symbols to upper/lower case: no, lower, upper).
**regexp** (expanded regular expressions [6] for the syntax check).
**name** (short description of the field).

The abbreviation of the record and the field name, keywords **bounds** and **color** should be presented with theirs values; other keywords may be absent.

2) The file COLORS.def describes the combination of colors of the text and the background corresponding to the index of the color from the file ENSDF.def. Next color fields should be defined:

text,
background,
misprint,
background for misprint,
marked text,
background for the marked text,
marked text with a misprint,
background for the marked text with a misprint.

Colors are defined by the hexagonal number #RRGGBB or by the word from the system file of colors /usr/X11/lib/X11/rgb.txt.

The version of EVE for the text terminals uses the file COLORS.def which has the other structure.

3) The file .eve.rc. This file determines attributes of the marked field or block. The variable mark_attribute determines the marked field attribute, the variable error_attribute determines an erroneous field attribute. Values of these variables are: blink, bold, reverse, underline (depending on the type of the terminal).

Variables save_mark_after_copy and save_mark_after_move determine whether the marked object should be marked after the execution of the corresponding command. Admissible values are: yes, no. For the version EVE for GTK+ variables common_font, helper, helpers, scheme_viewer may be used.

4) The file (~/.gnome/eve) describes variables common_font, helper and helpers. This file is used in the X Window environment. Priority for the setting up for these variables is: .eve.rc, ~/.gnome/eve, EVE defaults.

All the auxiliary files may be placed in the current directory (./), or in the user's home directory (~/), or in a service directory of the user (~/.eve/), or in a system directory (usually it is /usr/local/eve/). If the file .eve.rc is absent, values by default are used.

## Viewer for level schemes

The program data_set_viewer is the auxiliary tool and is developed for the graphical presentation of the level schemes of nuclei from the ENSDF datasets. The program gives a scaled picture of the scheme of any complexity. There are no viewers for the level schemes among the programs presented at the NNDC BNL site except for the ENSDAT code, which may be used for this aim but with strong limitations.

### *Start of the viewer*

The program may be started by the command:

data_set_viewer -i <ENSDF_file_name> ,

where filename should have the ENSDF format, or:

data_set_viewer --input-file=<ENSDF_file_name> .

If the option "-I" or "--input-file" is absent, the program waits for input of the ENSDF formatted file from the keyboard. It allows to redirect of the input from a file and may be useful, if the viewer is used in the integrated program package:

data_set_viewer <ENSDF_file_name

The program reads the input data set in the ENSDF format and displays the level scheme. Figure 2 presents the level scheme from the 14C 14B B- DECAY dataset.

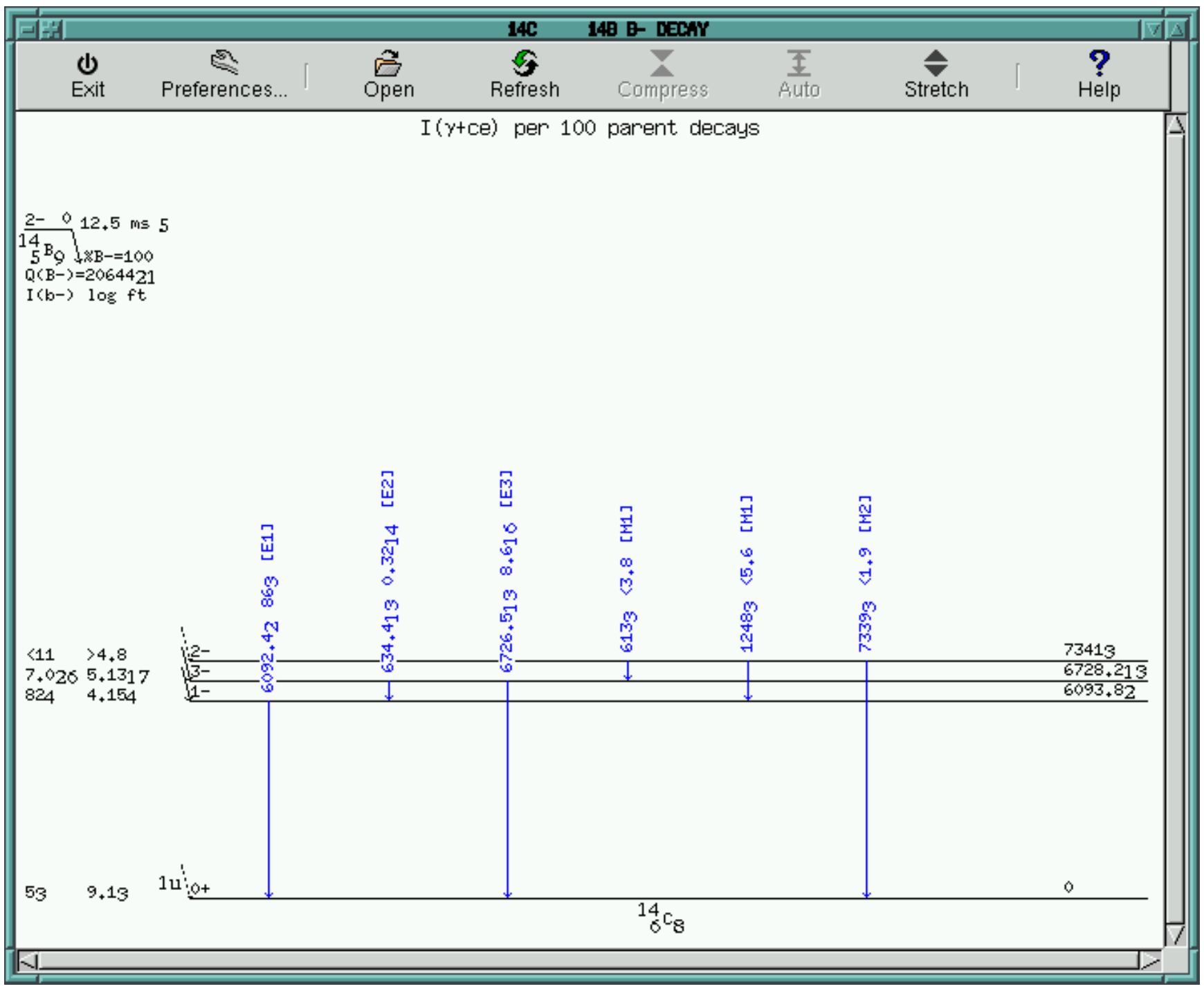

Fig. 2

## Panel of tools

### *Button "Exit"*

The program may be terminated by pressing the "Exit" button. The same result will be if the signal SIGTERM is sent to the program. It may be useful, if the viewer is used in the integrated program package.

### *Button "Open"*

The "Open" button calls the window for the input data file.

### *Button "Refresh"*

The "Refresh" button calls the re-reading of the data file. That makes it possible to edit the data file (with the EVE editor, e.g.) and at the same time to look through results not terminating the editor and other programs. The same result will be achieved, if the signal SIGHUP is sent to the data_set_viewer. It may be useful, if the program is used in the integrated program package.

### *Buttons "Compress", "Auto", "Stretch"*

These buttons are for change of scale along the vertical (energy) axis with the step of 2 respectively to the bottom of the image.

### *Button "Preferences"*

The "Preferences" button calls windows in which some options may be changed.

In the window "Font selection" 3 fonts may be chosen: the common font which is the used for values; a font for indexes which is used for values of errors (in units of the last digit) and a font for intensities of the gamma-radiation at the top of the panel.

In the window "Miscellaneous" uncertainties of the values may be switched on/off.

The viewer for the Help windows may be chosen in the window "Browser". "GNOME Less" and "less" are the text browsers. The "GNOME Help browser" works and looks like well known browsers "Netscape" and "Internet explorer".

The tuning of the viewer through the system of windows has some limitations. This procedure may be made more exact by the editing of the text configuration file data_set_viewer, which is situated in the ~/.gnome directory. Example:

```
 [Fonts]
common_font=-cronyx-fixed-medium-r-normal-*-10-*-*-*-c-*-koi8-r
index_font=-bitstream-charter-medium-r-normal-*-*-90-*-*-p-*-iso8859-9
greece_font=-etl-fixed-medium-r-normal-*-14-*-*-*-c-*-iso8859-7

[Misc]
draw_uncertainties=true
error_report=true
draw_gamma_values=true

[Help]
helper=3
helpers=less;;less\\ ;readme
 GNOME\\ Help\\ browser;DETACH+ANCHOR;gnome-help-browser\\ ;index.html
 Netscape;DETACH+ANCHOR;netscape\\ file://localhost;index.html
 w3m;DETACH+ANCHOR;w3m\\ -v\\ file://localhost;index.html
 lynx;ANCHOR;lynx\\ -restrictions=all\\ ;index.html
```

The work is supported by Russian Foundation for Basic Research, project № 09-07-00387-a.

Programs EVE and data_set_viewer, instructions for the installation and usage may be downloaded from the developer's site http://georg.pnpi.spb.ru. These programs are free distributed under GNU general public license as the non-commercial product. All remarks, proposals, additions will be accepted gratefully.

## References

1. B. Pritychenko, A.A. Sonzogni, D.F. Winchell, V.V. Zerkin, R. Arcilla, T.W. Burrows, C.L. Dunford, M.W. Herman, V. McLane, P. Oblozinsky, Y. Sanborn and J.K. Tuli. Nuclear reaction and structure data services of the National Nuclear Data Center, Annals of Nuclear Energy, Volume 33, Issue 4, March 2006, Elsevier, Pages 390-399.

2. The Nuclear Data Centres Network. IAEA Nuclear Data Section / Ed. V.G. Pronyaev. INDC(NDS)_401, IAEA, Vienna (1999).

3 W.B. Ewbank. Evaluated Nuclear Structure Data File (ENSDF) for Basic and Applied Research, p. 393 in Proceedings of the Fifth Biennial International CODATA conference, Boulder, Colorado, June 1976, ed. B. Dreyfus, Pergamon Press, Oxford and New York (1977).

4. J.K. Tuli. Evaluated Nuclear Structure Data File, A Manual for Preparation of Data Sets, BNL-NCS-51655-01/02-Rev Formal Report (February 2001).

5. G.I. Shulyak, A.A. Rodionov. The ENSDF_toolbox program package: tool for the evaluator of nuclear data, Report PNPI 2820 (2009), 20 p.

6. J. Peek, T. O'Reilly, M. Loukides. Unix Power Tools, O'Reilly & Associates Inc., Cambridge, Koln, Paris, Sebastopol, Tokyo (1997).